# Investigation of Random Laser in the Machine Learning Approach

**Emanuel P. Santos[1], Rodrigo F. Silva[2], Célio V. T. Maciel[2], Daniel F. Luz[2] and Pedro F. A. Silva[2]**

[1]Departamento de Física, Universidade Federal de Pernambuco, Recife-PE, Brazil
[2]Instituto de Física, Universidade Federal de Alagoas, Maceió-AL, Brazil
*E-mail: emanuel.pinheiro@ufpe.br

**Machine Learning and Deep Learning are computational tools that fall within the domain of artificial intelligence. In recent years, numerous research works have advanced the application of machine and deep learning in various fields, including optics and photonics. In this article, we employ machine learning algorithms to investigate the feasibility of predicting a stochastic phenomena: random laser emissions. Our results indicate that machine and deep learning have the capacity to accurately reproduce fluctuations characteristic of random lasers. By employing simple supervised learning algorithms, we demonstrate that the random laser intensity fluctuations can be predicted using spontaneous emission and pump intensity as input parameters in the models. Applications based on the demonstrated results are discussed.**

## 1. Introduction

Recently, some review articles about random lasers were published [1–3]. In these articles, the authors define random laser (RL) as laser-like emission in which the optical feedback is provided by scattering medium that combines a properly excited gain material and a random scattering structure. Additionally, they discuss about the advances and applications using random lasers like random fiber lasers, sensors, optical amplification and biomedical imaging. Despite the excellent discussion in these articles about the recent technological advances in the field of random lasers, the authors do not address works that employ Machine Learning (ML) or Deep Learning (DL) approaches to study random lasers.

Machine learning is a subfield of artificial intelligence that consists of computational techniques capable of learning from a dataset and then making predictions or estimations. Following that, deep learning is a subfield of machine learning that essentially refers to algorithms with neural network structures[4,5]. The ML has been extensively studied since it can be applied to social media features, product recommendations, image recognition, sentimental analysis, predict potential future data, language translation, among others. From an academic perspective, the ML attracts attention in finance [6], ecology [7], material science [8,9], physics[10], real world engineers applications [11] and others.

In context of optics and photonics, several works have explored the use of different algorithms of ML and DL for diverse perspectives and applications such as optical metrology [12], ultrafast photonics [13], photonic structure design [14,15], scanning near-field optical microscopy [16], luminescence thermometry [17,18], physics of lasers and fiber lasers [19,20] and others. However, some of these works have drawn our attention. Here[19,20] the authors show how to use some features of lasers to predict others. Particularly, is possible to use many characteristics as refractive index, wavelength, lamp current, pulse frequency, pulse width and others to predict features as effective index, dispersion, hole taper, etc. These works are promising because they bring the possibility that a set of characteristics serves to predict others sets that are totally different, but related. The works mentioned utilize ML and DL tools to study conventional lasers or materials in optics and photonics. To the best of our knowledge, there are no reported studies on this topic using ML techniques in the case of RL, perhaps because RL inherently exhibits stochasticity [21,22] in its emitted intensities.

Previous works has already shown that ML and DL algorithms may be able to make predictions and identify data characteristics even for stochastic processes [23–25]. Taking into consideration the computational power that ML can provide, we observe that even with simple algorithms, and despite the stochastic regime of intensity fluctuations of random lasers, machine learning can handle of these regimes and is able to reproduce the behavior of random laser emission.

Through the results, we discuss that applying ML to estimate the value of one part of the spectrum (RL) using another part of spectrum as an input parameter can be applicable in optical experiments, as proposed in the conclusion. Furthermore, another motivation for the study of this work is that the results presented demonstrate that ML can also be used as a technique to determine the threshold of laser action, distinguishing between regimes below and above the threshold.

## 2. Experiment and Algorithms

### 2.1. Experiment and Initial analyses

The experimental setup, as depicted in Figure SM1 of the supplementary material (SM), and the sample used ($Nd_{0.60}Y_{0.40}Al_3(BO_3)_4$) for the results in this article, are described in[26]. Under pulsed laser excitation at 532 nm and varying the excitation pulse energy (EPE) from 1.18 mJ to 6.39 mJ (involving 30 intervals), the random laser threshold was found at ~4.2 mJ. For each EPE we got 200 spectra some of which are shown in Figure 1, so we have 6000 spectra in total. Emissions were acquired by a spectrometer with a resolution of 2 nm, and the voltage of EPE was measured using an oscilloscope. The pump intensity at 532 nm was partially filtered so as not to damage the spectrometer. Fluorescent spontaneous emission has a broad band around 890 nm, corresponding to the $Nd^{3+}$ electronic transition from $^4F_{3/2}$ to $^4I_{9/2}$, whereas the electronic transition for the random laser is $^4F_{3/2}$ → $^4I_{11/2}$.

The right side of Figure 1 are the Pearson correlation coefficients for each pair among the measured quantities (excitation intensity at 532 nm, fluorescence intensity, random laser intensity and EPE) considering all data. The mathematical equation of this coefficient is:

$$C = \frac{\sum_{i=1}^{n}(x_i - \bar{x})(y_i - \bar{y})}{\sqrt{\sum_{i=1}^{n}(x_i - \bar{x})^2}\sqrt{\sum_{i=1}^{n}(y_i - \bar{y})^2}}, \quad (1)$$

where $\bar{x}$ and $\bar{y}$ are the averages of variables $x_i$ and $y_i$, respectively.

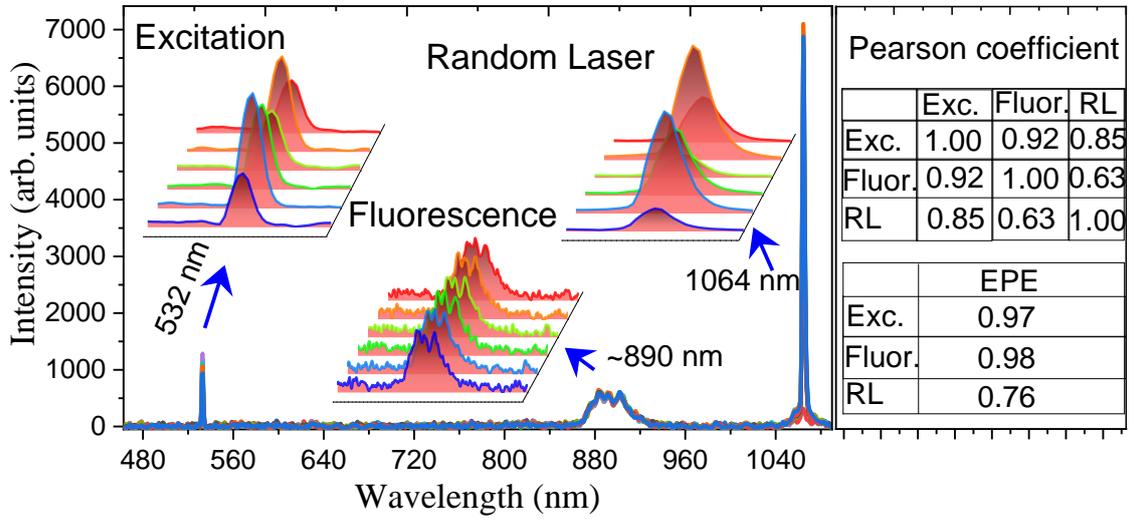

**Figure 1.** Full collected spectra from 463 nm to 1089 nm with EPE at 5.64 mJ. Here 532 nm refers to pump excitation, around 890 nm is fluorescence spontaneous emission and 1064 nm is random laser emission. The right side are the correlation coefficients between all possible combinations of the data set using all dataset.

As we can see, the correlation is greater between EPE and fluorescence (0.98) than between random laser and any other quantity shown in the Figure 1. However, there is a significant correlation between RL and excitation (0.85), which means that even when dealing with random emissions, it still carries information from the excitation pump. The lowest correlation occurs between RL and fluorescence (0.63), where this fact occurs due to the gain clamping [26], which shows that while RL grows rapidly, fluorescence can either increase or saturate. However, even with low correlation between these quantities, we can study the relations of fluorescence and excitation pulses with the random laser through ML algorithms, in other words, we can seek associations of a part of the spectrum (excitation + fluorescence) with another part of the spectrum (RL) through the ML (Figure 2).

2.2. Machine/Deep Learning algorithms

In the models we studied for our data, we employed supervised machine learning for regression – see a brief explanation about regression algorithms in SM – which involves providing labeled input data during the training [27]. We used fluorescent band data and excitation laser intensities (Figure 2) as input, labeled with 1064nm intensities, which also are the target variables (outputs) in the testing phase. Altogether, we have 6000 data (spectra), in which 75% were used in the training stage. We performed the training using 4 different algorithms, 3 of them machine learning (K Nearest Neighbors, ExtraTreesRegressor and StackingEstimator) and 1 deep learning (Convolutional Neural Network – CNN) in which we briefly describe them below.

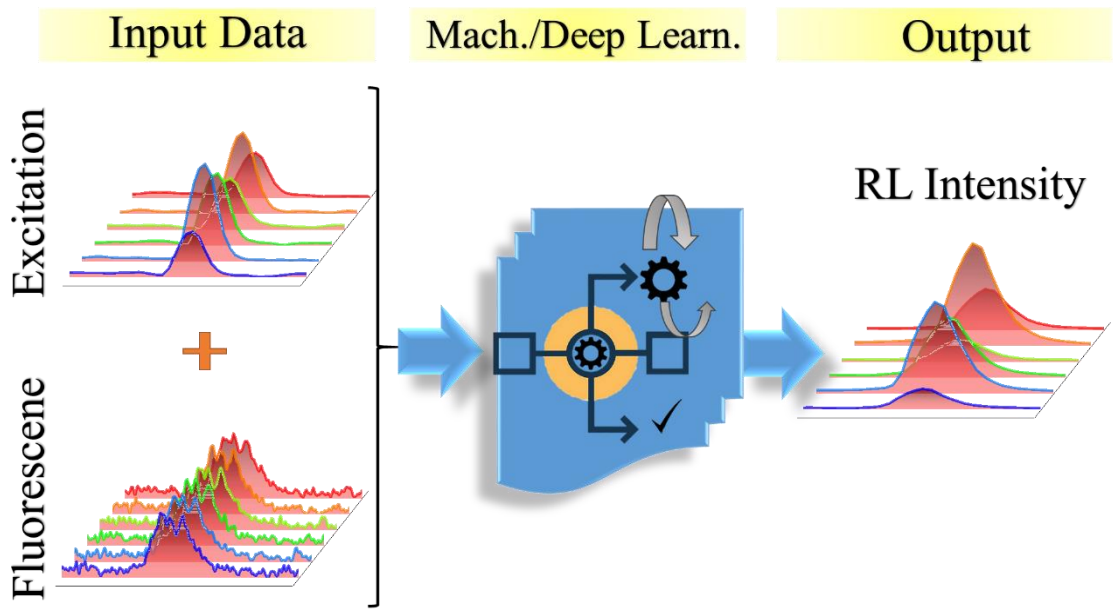

**Figure 2.** Supervised learning model to predict random laser emission (at 1064 nm) using as ML input parameters the spontaneous emission (at around 890 nm) and the pump intensity (at 532 nm). The machine/deep learning step is responsible to train, valid and test the data sets.

**K Nearest Neighbors (KNN):** The K Nearest Neighbors (KNN) method associates a certain quantity K of inputs (to be defined by the programmer) with a specific output parameter. Therefore, when new data is provided, the program checks which group of nearest neighbors this new data belongs to and makes predictions according to the parameters of that group. For a comprehensive description of the method, see [28].

**Decision Tree (ExtraTreesRegressor):** The ExtraTreesRegressor is an ensemble machine learning algorithm that utilizes decision trees to perform regression tasks. This algorithm operates by creating multiple independent decision trees in parallel. Each tree is trained on a random subset of features and a random sample of the training data. The predictions from individual trees are aggregated to make the final prediction. For a comprehensive description of the method, refer to[29].

**Convolutional Neural Network (CNN):** The convolutional neural network is a two-stage construction: i) preprocessing and ii) dense neural network. Within the preprocessing stages, we have convolution, which involves applying filters (or kernels) that act to remove irrelevant data, and we also have pooling layers that reduce the spatial dimension of features obtained through convolution, thereby reducing the number of parameters and accelerating the learning process, among other components. For a comprehensive description of the method, refer to[4].

**StackingEstimator:** StackingEstimator is a machine learning technique that enhances predictive accuracy by blending the outputs of multiple base models. In our approach, we leverage the TPOT library for automated model optimization, encompassing hyperparameter tuning and the selection of optimal base models. Specifically, our stacking configuration involves utilizing the ExtraTreesRegressor as the foundational estimator. Subsequently, we introduce the K Nearest Neighbors (KNN) algorithm to learn intricate combinations from the base models, leading to a refined and more accurate final prediction. For more details, consult[30].

For the data in this study, considering its characteristics and the provided amount of data, the stacking method demonstrated superior results, and the figures presented in this article are outcomes of this stacking approach. This model was also selected by automated machine learning (AutoML) – see brief description about AutoML in the SM – in order for us to determine this as the best model for our data. However, we tested all the aforementioned methods, and the comparison among them can be observed in table II and in the Figure SM3.

## 3. Results and Discussion

There are many ways to highlight the action regime of the random laser during the threshold, such as spectral narrowing[31,32], the efficiency slope of input/output intensities[33–35], decay time shortening[36], the $\alpha$ parameter of the Lévy distribution[37,38], the Parisi overlap parameter[39] and more recently, the relationship between spontaneous emission and stimulated emission[26]. Among these methods, we will consider the $\alpha$ parameter of the $\alpha$-stable Lévy distribution [40] $P(t)$ to characterize the intensity fluctuations (represented by $t$ in the equation below) of the RL:

$$P(t) = \exp\{it\mu - |ct|^\alpha[1 - i\beta sgn(t)\Phi]\} \qquad (2)$$

with

$$\Phi = \begin{cases} \tan\left[\frac{\alpha\pi}{2}\right], & \alpha \neq 1 \\ -(2/\pi)\ln|t|, & \alpha = 1 \end{cases}. \tag{3}$$

In these equations, $= \sqrt{-1}$, $\mu \in \mathbb{R}$, $c > 0$, $\beta \in [-1,1]$ and $\alpha \in (0,2]$ is the Lévy index. The Lévy index $0 < \alpha < 2$ classifies the intensity fluctuations of random laser around the threshold, whereas for $\alpha = 2$, it characterizes Gaussian behaviors that appears below or well above the threshold [38,41].

To demonstrate the behavior of random laser intensity fluctuations, we present in Figure 3(a) the normalized intensities at 1064 nm during the acquisition of 200 sepctra for four different EPE. For an energy below the threshold (at 1.18 mJ), the intensity points fluctuate following a Gaussian distribution, i.e., $\alpha = 2$. Around and near the threshold (4.35 mJ and 4.86 mJ), the intensities fluctuate according to the Lévy flight for light[42,43] ($0 \leq \alpha \leq 2$). However, the Gaussian regime tends to be reestablished for higher energies. Using the modeling described in section 2.2, we obtain the predicted intensities (Figure 3b) from the algorithm and compare them with the measurements (Figure 3a). Additionally, histograms were displayed for each indicated EPE (Figure 3c).

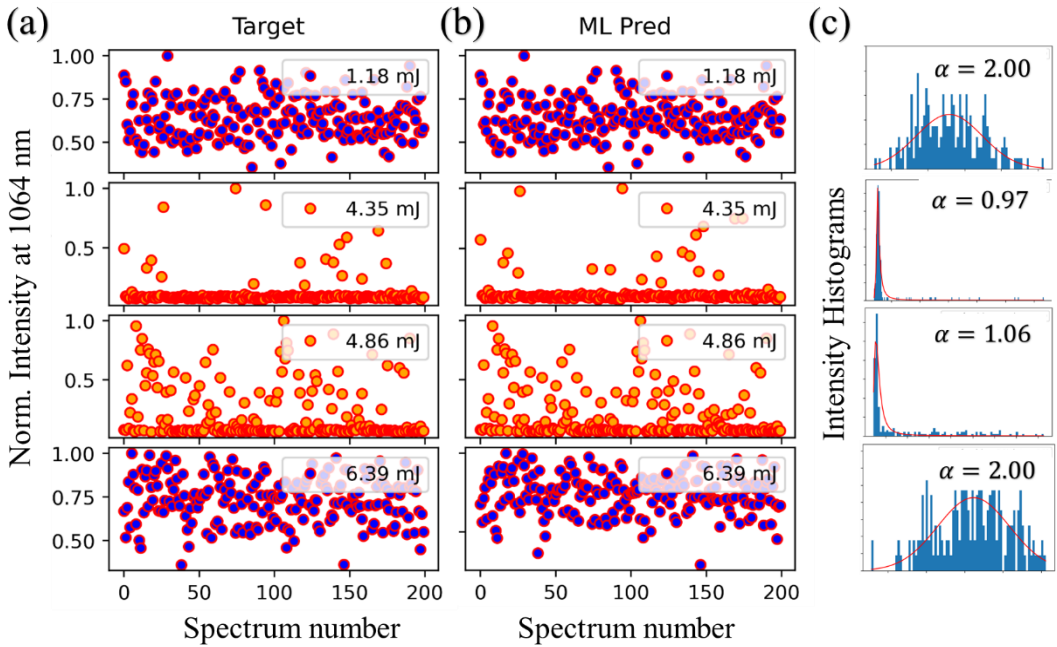

Figure 3. Fluctuations in the (a) target and (b) predicted RL intensity at a fixed EPE. (c) Histograms with Lévy index obtained through fitting (solid red line) using equation 2.

The choice of fluorescence, spontaneous emission and/or excitation pulse energy as input parameters for making predictions was made by observing the correlations shown in Figure 1. For machine learning, it is important to consider input parameters that are in some way correlated with the output data. Although random laser emissions exhibit fluctuations, the entire spectrum is correlated, and therefore, using a portion of the spectrum becomes a good set of input data.

The stochastic regime of intensity fluctuations of random lasers has led the study of the RLs into the statistical physics of complex systems, enabling experimental observations of various statistical phenomena such as the Lévy flights, replica symmetry breaking, spin glass behavior of light and turbulence[44], for instance. Although well explored, random lasers have not been studied in the domain of machine or deep learning.

Considering now all 6000 spectra, with the excitation pulse energy fixed every 200 acquisitions, we display the intensities at 1064 nm for each of the spectra, along with the average value every 200 intervals, and the predictions made by the algorithm in Figure 4.

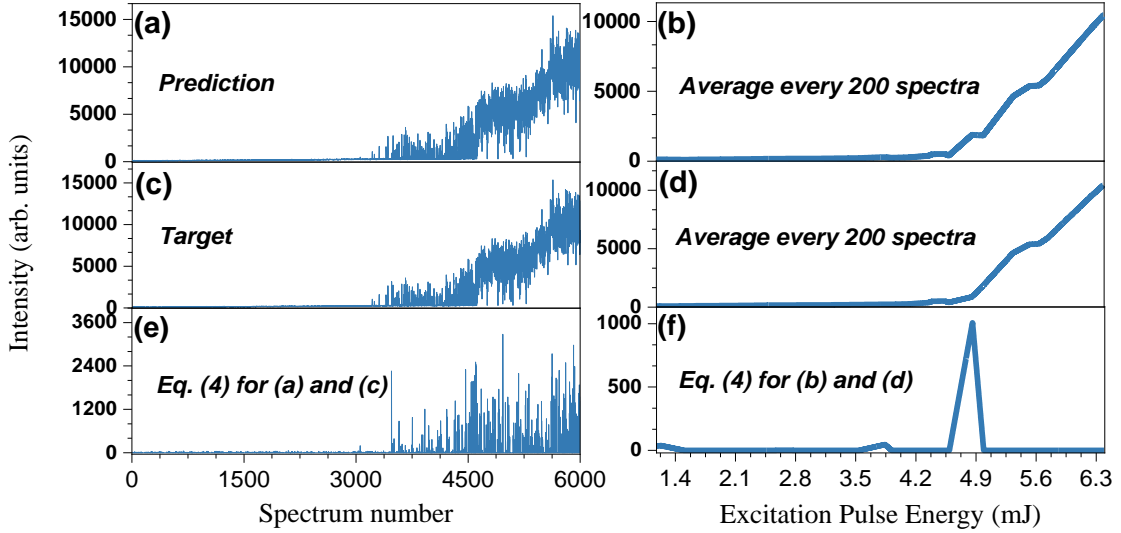

Figure 4. Prediction (a-b) vs Target (c-d) intensities of RL at 1064 nm and the difference between prediction and target (e-f); (a) all predicted intensities and (b) prediction of average every 200 spectra, corresponding to a given fixed EPE; (c) measured intensities (target) and (d) the corresponding averages; (e-f) absolute difference between predicted and target using.

At first glance from Figure 4(a-d), it may appear that the predictions produce behavior equal to the experiment. However, what actually occurs is that ML is capable of reproducing the behavior, but it cannot precisely match the peak locations. In other words, many peaks predicted by the algorithm are slightly shifted compared to the experimental positions. Nevertheless, for clearer visualization, Figure 4(e-f) illustrates the difference between target and prediction variables using the following equation:

$$\sqrt{(y_i - x_i)^2}, \qquad (4)$$

where $y_i$ is the predicted variable of the $x_i$ target. Notably, the prediction uncertainty increases in the region from the threshold, which we attribute to the intrinsic stochasticity of the random laser process. Also, another factor is the relatively small amount of data, since ML algorithms are more effective with a large amount of data. However, for the proof of concept, the quantity used in this work is sufficient.

To evaluate the ML models for the dataset, we used the following metric parameters:

- Mean Absolute Error (MAE):

$$MAE = \frac{1}{n}\sum_{i=1}^{n} |y_i - \hat{y}_i|. \quad (5)$$

- Root Mean Squared Error (RMSE):

$$RMSE = \sqrt{\frac{1}{n}\sum_{i=1}^{n} (y_i - \hat{y}_i)^2}. \quad (6)$$

- Maximum Error (ME).

Where $y_i$ is the experimental (target) value, and $\hat{y}_i$ is the predicted value. And the maximum error is the highest value in the list calculated by equation 4.

The metrics were calculated by comparing the predictions to the targets for the full dataset, which includes both training and testing data, as well as for the full test set, the subset of the test set comprising intensities below the threshold, and the subset above the threshold. These results were organized in Table I.

Table I. Model evaluation for the full dataset, full test set, points below and above the threshold.

|      | Full dataset | Full test set | Subset of test set / Below the threshold | Subset of test set / Above the threshold |
|------|---|---|---|---|
| **RMSE** | 217.15 | 560.68 | 6.87 | 369.22 |
| **ME** | 3276.40 | 3276.40 | 55.34 | 3276.40 |
| **MAE** | 38.31 | 255.37 | 2.15 | 105.57 |

Considering the full test set, we compared the models (Figure SM3) described in section 2.2, as shown in Table II.

Table II – Evaluation of the full test set for the 4 algorithms described in section 2.2.

|      | KNN | Extra Trees | CNN | Stacking Estimator |
|------|---|---|---|---|
| **RMSE** | 1055.52 | 583.89 | 577.47 | 560.68 |
| **ME** | 5515.75 | 3298.37 | 2671.73 | 3276.40 |
| **MAE** | 523.42 | 282.66 | 278.36 | 255.37 |

Since these metric parameters essentially measure how good the model is for the given amount of data, the smaller the value of these metrics, the better the model is at making prediction.

In general, the use of artificial intelligence models involves continuous training so that the machine can learn from an ever-expanding dataset. The models presented in this work can also be further improved and adapted by considering the use of a spectrometer with better temporal resolution, a larger amount of acquired data, a greater number of input parameters, or even a very extensive dataset of experiments from different random laser emitters that. Of course, many tests can be conducted to determine the best prediction, such as, for example, what input data are required and how much is needed, and what are the best features to use. Among various tests, the configurations we have presented have proven to be the most effective for our data. Additionally, the data can be tested with more complex machine and deep learning algorithms such as Convolutional Neural Networks, Kernel Support Vector Machine, Deep Bayesian Networks, Generative Adversarial Networks, etc.

## 4. Conclusion

In this study, we establish a connection between photonics, statistical mechanics, and data science, demonstrating that ML holds significant potential for assessing the behavior of random lasers. This potential is evident even when utilizing straightforward algorithms that do not demand extensive computational resources.

The advancement of this study paves the way for discussions on how modern computational techniques can be employed within the realm of optics and photonics of complex systems. We demonstrate that the behavior observed in one part of the spectrum can suffice to deduce occurrences in another part of the spectrum. In this research, the machine was able to replicate events at 1064 nm by solely considering the bands around 900 nm and the excitation pump at 532 nm.

A simple application example: imagine in your laboratory, you have two spectrometers, where only one can measure at 1064 nm, while the second one can only measure below 1000 nm. Suppose the first spectrometer is either broken or in use, and you need to use the second one. Even though you can only view data up to 1000 nm on the screen, since the algorithm is already trained to predict what happens at 1064 nm, the experimenter can determine whether there is a random laser (RL) or not. Thus, using the method presented in this study, you simply need to train the algorithm using one part of the spectrum to attempt to predict events in other parts. Of course, this example can be extended to similar situations.

Another application that we highlight in this work is that machine learning can be used to determine the operating threshold of random lasers. For example, based on Figure 4 and Table 1, it can be concluded that there is a significant difference between predictions before and after the threshold, thus positioning machine learning as an effective threshold detector for random lasers.

In conclusion, we have demonstrated that random lasers provide an interesting platform for studying machine learning algorithms, even if they are simple algorithms. The results can also be viewed as a didactic approach to studying data with fluctuations and extracting relevant information from it.

**Data availability statement**

The raw data are available upon reasonable request from the authors. The code and more results are in the supplementary material.

**Conflict of Interest**

The authors declare no conflict of interest.